\documentclass[submission,copyright,creativecommons]{eptcs}
\providecommand{\event}{WPTE 2016} 
\usepackage{breakurl,latexsym,amssymb,proof,graphics,color} 

\title{On Upper Bounds on the Church-Rosser Theorem}
\author{Ken-etsu Fujita
\institute{Department of Computer Science\\
Gunma University\\
Kiryu, Japan}
\email{fujita@cs.gunma-u.ac.jp}
}
\def\titlerunning{On Upper Bounds on the Church-Rosser Theorem}
\def\authorrunning{K. Fujita}

\newtheorem{definition}{Definition}
\newtheorem{lemma}{Lemma}
\newtheorem{theorem}{Theorem}
\newtheorem{proposition}{Proposition}
\newtheorem{corollary}{Corollary}

\newtheorem{example}{Example}

\begin{document}
\maketitle


\begin{abstract}
The Church-Rosser theorem in the type-free $\lambda$-calculus
is well investigated both  for 
$\beta$-equality and $\beta$-reduction. 
We provide a new proof of the theorem for $\beta$-equality 
with no use of parallel reductions, but simply with Takahashi's translation 
(Gross-Knuth strategy). Based on this,  
upper bounds for reduction sequences on the theorem are obtained as 
the fourth level of the Grzegorczyk hierarchy.
\end{abstract}

\section{Introduction} 

\subsection{Background}

The Church-Rosser theorem \cite{CR1936} is one of the most fundamental 
properties on rewriting systems, which guarantees uniqueness of computation 
and consistency of a formal system. 
For instance, 
for proof trees and formulae of logic
the unique normal forms of the corresponding terms and types in
a Pure Type System (PTS) can be chosen as their denotations \cite{TF1992}
via the Curry-Howard isomorphism.

The Church-Rosser theorem for $\beta$-reduction states that 
if $M\twoheadrightarrow N_1$ and $M\twoheadrightarrow N_2$
then we have $N_1\twoheadrightarrow P$ and $N_2\twoheadrightarrow P$
for some $P$. Here, we write $\twoheadrightarrow$ for the reflexive 
and transitive closure of one-step reduction $\to$. 
Two proof techniques of the theorem are well known;  
tracing the residuals of redexes along a sequence of reductions 
\cite{CR1936,Barendregt1984,HS2008},  
and working with parallel reduction 
\cite{CFC1974,Barendregt1984,HS2008,Takahashi1989} 
known as the method of Tait and Martin-L\"of. 
Moreover, a simpler proof of the theorem is established only with 
Takahashi's translation \cite{Takahashi1989}  
(the Gross-Knuth reduction strategy \cite{Barendregt1984}),  
but with no use of parallel reduction \cite{Loader1998,DVO2008}. 

On the other hand, 
the Church-Rosser theorem for $\beta$-equality states that 
if $M =_\beta N$ then there exists $P$ such that 
$M\twoheadrightarrow P$ and $N\twoheadrightarrow P$. 
Here, we write $M =_\beta N$ iff $M$ is obtained from $N$ 
by a finite series of reductions $(\twoheadrightarrow)$ and 
reversed reductions $(\twoheadleftarrow)$.
As the Church-Rosser theorem for $\beta$-reduction has been well studied, 
to the best of our knowledge 
the Church-Rosser theorem for $\beta$-equality is always {\it secondary} proved 
as a corollary from the theorem for 
$\beta$-reduction \cite{CR1936,CFC1974,Barendregt1984,HS2008}. 

One of our motivations is to analyze quantitative properties in general 
of reduction systems. 
For instance, measures for developments are investigated by  
Hindley \cite{Hindley1978} and de Vrijer \cite{Vrijer1985}.
Statman \cite{Statman1979} proved that deciding the $\beta\eta$-equality 
of typable $\lambda$-terms is not elementary recursive.
Schwichtenberg \cite{Schwichtenberg1982} analysed 
the complexity of normalization in the simply typed lambda-calculus, 
and showed that the number of reduction steps necessary to reach the 
normal form is bounded by a function at the fourth level of the 
Grzegorczyk hierarchy $\varepsilon^4$ \cite{Grzegorczyk1953}, 
i.e., 
a non-elementary recursive function. 
Later Beckmann \cite{Beckmann2001} determined the exact bounds for 
the reduction length of a term in the simply typed $\lambda$-calculus.  
Xi \cite{Xi1999} showed bounds for the number of 
reduction steps on the standardization theorem, and its application to 
normalization. 
In addition, Ketema and Simonsen \cite{KS2013} extensively studied 
valley sizes of confluence and the Church-Rosser property in 
term rewriting and $\lambda$-calculus as a function of given term sizes and 
reduction lengths. 
However, there are no known bounds for the Church-Rosser 
theorem for $\beta$-equality. 

In this study, 
we are also interested in quantitative analysis of the witness of the 
Church-Rosser theorem: 
how to find common contractums with the least size 
and with the least number of reduction steps.
For the theorem for $\beta$-equality 
($M=_\beta N$ implies $M\twoheadrightarrow^{l_3} P$ 
and $N\twoheadrightarrow^{l_4} P$ for some $P$), 
we study functions that set bounds on 
the least size of a common contractum $P$, 
and 
the least number of reduction steps $l_3$ and $l_4$ 
required to arrive at a common contractum, 
involving the term sizes of $M$ and $N$, and the length of $=_\beta$.
For the theorem for $\beta$-reduction 
($M\twoheadrightarrow^{l_1} N_1$ and $M\twoheadrightarrow^{l_2} N_2$ implies 
$N_1\twoheadrightarrow^{l_3} P$ and $N_2\twoheadrightarrow^{l_4} P$ 
for some $P$), 
we study functions that set bounds on 
the least size of a common contractum $P$, and 
the least number of reduction steps $l_3$ and $l_4$ 
required to arrive at a common contractum, 
involving the term size of $M$ and the lengths of $l_1$ and $l_2$.

\subsection{New results of this paper}

In this paper, first we investigate {\it directly} 
the Church-Rosser theorem for 
$\beta$-equality 
{\it constructively} 
from the viewpoint of 
Takahashi translation \cite{Takahashi1989}. 
Although the two statements are equivalent to each other, 
the theorem for $\beta$-reduction is a special case of that for 
$\beta$-equality. 
Our investigation shows that a common contractum of $M$ and $N$ such that 
$M =_\beta N$ 
is determined 
by (i) 
$M$ and the number of occurrences of reduction $(\to)$ appeared in $=_\beta$,  
and also 
by (ii) $N$ and that of reversed reduction $(\leftarrow)$. 
The main lemma plays a key role 
and reveals a new invariant 
involved in the equality $=_\beta$, independently of 
an exponential combination of reduction and reversed reduction. 
Next, in terms of iteration of translations,  
this characterization of the Church-Rosser theorem
makes it possible to analyse how large common contractums are 
and how many reduction-steps are required to 
obtain them. 
From this, 
we obtain an upper bound function for the theorem in    
the fourth level of the Grzegorczyk hierarchy. 
In addition, the theorem for $\beta$-reduction is handled 
as a {\it special case} of the theorem for $\beta$-equality, 
where the key notion is contracting new redexes under development.

\subsection{Outline of paper}

This paper is organized as follows. 
Section 1 is devoted to background, related work, and new results 
of this paper. 
Section 2 gives preliminaries including basic definitions and notions.
Following the main lemma,  
Section 3 provides a new proof of the Church-Rosser theorem 
for $\beta$-equality. 
Based on this, reduction length and term size for the theorem 
are analyzed in Section 4, and then we compare with related results.
Section 5 concludes with remarks, related work, and further work.

\section{Preliminaries}

The set of $\lambda$-terms denoted by $\Lambda$ is defined 
with a countable set of variables as follows.
\begin{definition}[$\lambda$-terms]
$$
M, N, P, Q\in \Lambda ::= x \mid (\lambda x. M) \mid (MN)
$$
\end{definition}
We write $M\equiv N$ for the syntactical identity under renaming 
of bound variables. We suppose that every bound variable is distinct from
free variables. The set of free variables in $M$ is denoted by $\mathrm{FV}(M)$.

If $M$ is a subterm of $N$ then we write $M\sqsubseteq N$ for this.
In particular, we write $M\sqsubset N$ if $M$ is a proper subterm of $N$.
If $P\sqsubseteq M$ and $Q\sqsubseteq M$, and moreover 
there exist no terms $N$ such that $N\sqsubseteq P$ and $N\sqsubseteq Q$, 
then we write $P\parallel Q$ for this, i.e., 
$P$ and $Q$ have non-overlapping parts of $M$.  
\begin{definition}[$\beta$-reduction]
One step $\beta$-reduction $\to$ is defined as follows, where 
$M[x:=N]$ denotes a result of substituting $N$ for every free occurrence of 
$x$ in $M$. 
\begin{enumerate}
\item $(\lambda x.M)N \to M[x:=N]$

\item If $M\to N$ then $PM \to PN$, $MP \to MP$, and 
$\lambda x.M \to \lambda x.N$.
\end{enumerate}
\end{definition}
A term of the form of $(\lambda x.P)Q\sqsubseteq M$ is called a redex of $M$. 
A redex is denoted by $R$ or $S$, and we write $R: M \to N$ 
if $N$ is obtained from $M$ by contracting the redex $R\sqsubseteq M$.
We write $\twoheadrightarrow$ for the reflexive and transitive closure 
of $\to$. 
If $R_1: M_0 \to M_1, \ldots, R_{n}: M_{n-1}\to M_n$ $(n\geq 0)$, 
then for this we write $R_0\ldots R_n: M_0 \twoheadrightarrow^n M_n$, 
and the {\it reduction sequence} is denoted by the list 
$[M_0, M_1, \ldots, M_n]$. 
For operating on a list, 
we suppose fundamental list functions, {\tt append}, {\tt reverse}, 
and {\tt tail} ({\tt cdr}).
\begin{definition}[$\beta$-equality]
A term $M$ is $\beta$-equal to $N$ with reduction sequence $ls$, 
denoted by $M =_\beta N$ with $ls$ is defined as follows: 
\begin{enumerate}
\item If $M\twoheadrightarrow N$ with reduction sequence $ls$, 
then $M =_\beta N$ with $ls$.
\item If $M =_\beta N$ with $ls$, then $N=_\beta M$ with ${\tt reverse}(ls)$.
\item If $M =_\beta P$ with $ls_1$ and $P=_\beta N$ with $ls_2$, then 
$M =_\beta N$ with ${\tt append}(ls_1, {\tt tail}(ls_2))$. 
\end{enumerate}
\end{definition}
Note that 
$M =_\beta N$ with reduction sequence $ls$ 
iff 
there exist terms $M_0, \ldots, M_n (n\geq 0)$ in this order 
such that $ls = [M_0, \ldots, M_n]$, 
$M_0\equiv M, M_n\equiv N$, and 
either $M_i \to M_{i+1}$ or $M_{i+1}\to M_i$ for each $0\leq i\leq n-1$.   
In this case, we say that 
the {\it length} of $=_\beta$ is $n$, denoted by $=_\beta^n$. 
The arrow in $M_i \to M_{i+1}$ is called a {\it right arrow}, 
and the arrow in $M_{i+1}\to M_i$ is called a {\it left arrow},
denoted also by $M_i \leftarrow M_{i+1}$. 
\begin{definition}[Term size]
Define a function $\vert~ \vert : \Lambda \to {\mathbf N}$ as follows.
\begin{enumerate}
\item $\vert x\vert =1$

\item $\vert \lambda x.M\vert = 1+\vert M\vert$

\item $\vert M N \vert = 1 + \vert M\vert + \vert N\vert$
\end{enumerate}
\end{definition}
\begin{definition}[Takahashi's * and iteration]
The notion of Takahashi translation $M^*$ \cite{Takahashi1989}, that is, 
the Gross-Knuth reduction strategy \cite{Barendregt1984} 
is defined as follows.  
\begin{enumerate}
\item $x^* = x$

\item $((\lambda x.M)N)^* = M^*[x:= N^*]$

\item $(MN)^* = M^* N^*$

\item $(\lambda x.M)^* = \lambda x.M^*$
\end{enumerate}
The 3rd case above is available provided that $M$ is not in the form of 
a $\lambda$-abstraction.
We write an iteration of the translation \cite{Takahashi1991} as follows. 
\begin{enumerate}
\item $M^{0*} = M$

\item  $M^{n*} = (M^{(n-1)*})^*$
\end{enumerate}
\end{definition}
We write $\sharp (x\in M)$ for 
the number of free occurrences of the variable $x$ in $M$.
\begin{lemma}
$\vert M[x:= N] \vert = \vert M\vert + \sharp(x\in M)\times(\vert N\vert -1)$.
\end{lemma}
{\it Proof.} By straightforward induction on $M$. 
\hfill $\Box$
\begin{definition}[$\mathsf{Redex}(M)$]
The set of all redex occurrences in a term $M$ is 
denoted by $\mathsf{Redex}(M)$. 
The cardinality of the set $\mathsf{Redex}(M)$ is denoted by 
$\sharp \mathsf{Redex}(M)$.
\end{definition}
\begin{lemma}[$\sharp\mathsf{Redex}(M)$]\label{lemma:redex}
We have $\sharp\mathsf{Redex}(M) \leq \frac{1}{2}\vert M\vert -1$
for $\vert M\vert \geq 4$. 
\end{lemma}
{\it Proof.} 
Note that $\sharp\mathsf{Redex}(M) =0$ for $\vert M\vert < 4$. 
By straightforward induction on $M$ for $\vert M\vert \geq 4$. 
\hfill $\Box$
\begin{lemma}[Substitution]\label{lemma:substitution}
If $M_1 \twoheadrightarrow^{l_1} N_1$ 
and $M_2\twoheadrightarrow^{l_2} N_2$,  
then $M_1[x:= M_2] \twoheadrightarrow^{l} N_1[x:= N_2]$ 
where $l =  l_1 + \sharp(x\in M_1) \times l_2$. 
\end{lemma}
{\it Proof.} 
By induction on the derivation of $M_1\twoheadrightarrow^{l_1} N_1$.
The case of $l_1=0$ requires induction on $M_1\equiv N_1$. 
We also need induction on the derivation of $M_1 \to N_1$, and 
we show here one of the interesting cases.
\begin{enumerate}
\item Case of $(\lambda y.M)N \twoheadrightarrow^1 M[y:=N]$: 
\begin{eqnarray*}
(\lambda y.M[x:= M_2])(N[x:= M_2])
 &\twoheadrightarrow^{m_1} & (\lambda y.M[x:= N_2])(N[x:= M_2])
\mbox{ by IH1} 
\\
 &\twoheadrightarrow^{m_2} & (\lambda y.M[x:= N_2])(N[x:= N_2])
\mbox{ by IH2} 
\\
 &\twoheadrightarrow^{1} & (M[x:= N_2])[y:= (N[x:= N_2])] 
\end{eqnarray*}
Here, IH1 is $\lambda y.M[x:= M_2] 
\twoheadrightarrow^{m_1} \lambda y.M[x:= N_2]$ 
with $m_1 = \sharp(x\in M)\times l_2$. 
IH2 is $N[x:= M_2] 
\twoheadrightarrow^{m_2} N[x:= N_2]$ with $m_2 = \sharp(x\in N)\times l_2$. 
Therefore, 
\begin{eqnarray*}
l &=& m_1 + m_2 + 1
\\
&=& 1 + \sharp(x\in M)\times l_2 + \sharp(x\in N)\times l_2 
\\
&=& 1 + \sharp(x\in ((\lambda y.M)N))\times l_2.
\hspace{93mm} \Box
\end{eqnarray*}
\end{enumerate}
\begin{proposition}[Term size after $n$-step reduction]\label{prop:length}
If $M\twoheadrightarrow^n N$ $(n\geq 1)$ 
then 
\[\vert N\vert 
< 8
\left(
\displaystyle{\frac{\vert M\vert}{8}}
\right)^{2^n}
.\]
\end{proposition}
{\it Proof.}  By induction on $n$. 
\begin{enumerate}
\item Case of $n=1$, where $M\to M_1$:

The following inequality can be proved 
by induction on the derivation of $M\to M_1$:
\[
\vert M_1\vert \leq  \frac{\vert M\vert^2}{2^3} -1
\] 
\item Case of $n=k+1$, where $M\to M_1\twoheadrightarrow^{k} M_{k+1}$: 
\begin{eqnarray*}
\vert M_{k+1}\vert &<& 8\left({\frac{\vert M_1\vert}{8}}\right)^{2^k}
 \mbox{ ~ ~ ~ from the induction hypothesis}
\\
 &<& 8\left(\left(\frac{\vert M\vert}{8}\right)^2\right)^{2^k}
 \mbox{ from $\vert M_1\vert < \frac{1}{8}\vert M\vert^2$}
\\
& =& 8\left(\frac{\vert M\vert}{8}\right)^{2^{(k+1)}} 
\hspace{103mm} \Box
\end{eqnarray*}
\end{enumerate}
\begin{lemma}[Size of $M^*$]\label{lemma:sizeofstar}
We have $\displaystyle{\vert M^*\vert \leq 2^{\vert M\vert-1}}$.
\end{lemma}
{\it Proof.}  
By straightforward induction on $M$.  \hfill $\Box$
\begin{definition}[Residuals \cite{CR1936,HS2008}]\label{def:residuals}
Let $\mathcal{R} \subseteq \mathsf{Redex}(M)$. 
Let $R\in \mathcal{R}$, and $R: M\to N$. 
Then the set of residuals of $\mathcal{R}$ in $N$ with respect to $R$, 
denoted by 
$\mathsf{Res}(\mathcal{R}/ R: M\to N)$ 
is defined by the smallest set satisfying the following conditions:
\begin{enumerate}
\item Case of $S\in\mathcal{R}$ and $S\parallel R$: 

Then we have $S\in \mathsf{Res}(\mathcal{R}/ R: M\to N)$.

\item Case of $S\in\mathcal{R}$ and $S\equiv R$:  

Then we have $S\not\in \mathsf{Res}(\mathcal{R}/ R: M\to N)$.

\item Case of $S\in\mathcal{R}$ 
and $S\equiv (\lambda x.M_1) N_1$
and $R \sqsubset M_1$ for some $M_1,N_1\sqsubset M$:

Then we have $S'\in \mathsf{Res}(\mathcal{R}/ R: M\to N)$
such that $R: S\to S'$ for $S'\sqsubset N$.

\item Case of $S\in\mathcal{R}$ and 
$S\equiv (\lambda x.M_1) N_1$
and $R \sqsubset N_1$ for some $M_1,N_1\sqsubset M$:

Then we have $S'\in \mathsf{Res}(\mathcal{R}/ R: M\to N)$
such that $R: S\to S'$ for $S'\sqsubset N$.

\item Case of $S\in\mathcal{R}$ and $R\equiv (\lambda x.M_1)N_1$ 
and $S\sqsubset M_1$ for some $M_1,N_1\sqsubset M$: 

Then we have $S[x:= N_1]\in \mathsf{Res}(\mathcal{R}/ R: M\to N)$
such that $S[x:= N_1] \sqsubset M_1[x:= N_1]$ where 
$R: (\lambda x.M_1)N_1 \to M_1[x:= N_1]$. 

\item Case of $S\in\mathcal{R}$ and 
$R\equiv (\lambda x.M_1)N_1$ and 
$S\sqsubset N_1$ for some $M_1,N_1\sqsubset M$: 

Then we have $S\in \mathsf{Res}(\mathcal{R}/ R: M\to N)$ 
for every occurrence $S$ such that 
$S\sqsubset M_1[x:= N_1]$ where 
$R: (\lambda x.M_1)N_1 \to M_1[x:= N_1]$. 
\end{enumerate}
\end{definition}
\begin{definition}[Complete development \cite{Barendregt1984}]\label{def:development}
Let $\mathcal{R}\subseteq \mathsf{Redex}(M)$. 
A reduction path $R_0 R_1\ldots: M\equiv M_0\to  M_1 \to \cdots $ 
is a development of $\langle M, \mathcal{R}\rangle$ if and only if   
each redex $R_i\sqsubseteq M_i$ is in the set $\mathcal{R}_i$  
$(i\geq 0)$ such that $\mathcal{R}_0 = \mathcal{R}$ and 
$\mathcal{R}_i = \mathsf{Res}(\mathcal{R}_{i-1}/R_{i-1}: M_{i-1}\to M_i)$.
If $\mathcal{R}_k = \emptyset$ for some $k$, then 
the development is called complete. 
\end{definition}
\begin{definition}[Minimal complete development \cite{HS2008}]\label{def:minimal}
Let $\mathcal{R}\subseteq \mathsf{Redex}(M)$. A redex occurrence 
$R\in \mathcal{R}$ is called minimal if 
there is no $S\in\mathcal{R}$ such that $S\sqsubset R$
(i.e., $R$ properly contains no other $S\in\mathcal{R}$).

Let $\mathcal{R} = \{R_0, \ldots, R_{n-1}\}$.  
Let $\mathcal{R}_0 = \mathcal{R}$
and $\mathcal{R}_i = \mathsf{Res}(\mathcal{R}_{i-1}/R_{i-1})$. 
A reduction path $M \twoheadrightarrow^n N$ is 
a minimal complete development of $\mathcal{R}$ 
if and only if 
we contract any minimal $R_i\in \mathcal{R}_i$ 
at each reduction step.  
This development is also called an inside-out development
that yields shortest complete developments 
\cite{Khasidashvili1988,Sorensen2007}. 

We write $M \Rightarrow N$ 
if $N$ is obtained from $M$ by a minimal complete development 
of a subset $\{R_1, \ldots, R_n\}$ of $\mathsf{Redex}(M)$. 
In this case, we write $R_1\ldots R_{n}: M\Rightarrow^n N$. 
\end{definition}
Note that we can repeat this development at most $n$-times 
with respect to $\mathcal{R}= \{R_0, \cdots, R_{n-1}\}$ 
until no residuals of $\mathcal{R}$ are left, 
since we never have the fifth or sixth case in Definition \ref{def:residuals},
and then we have $R\not\in\mathsf{Res}(\mathcal{R}/R)$. 
\begin{definition}[Reduction of new redexes]\label{def:new-redex}
Let $R\!:\!M\!\!\to\!\!N$. 
If there exists a redex occurrence $S\!\!\in\!\mathsf{Redex}(N)$ 
but $S\not\in\mathsf{Res}(\mathsf{Redex}(M)/R: M\to N)$, 
then we say that 
the reduction $R: M\to N$ creates a new redex $S\sqsubseteq N$, and 
$N$ contains a created redex after contracting $R$.

Let $\sigma$ be  
a reduction path $R_0 R_1\ldots: M\equiv M_0\to  M_1 \to \cdots $. 
We define the set of new redex occurrences denoted by 
$\mathsf{NewRed}(M_{i+1})$ $(i\geq 0)$ as follows: 
\begin{eqnarray*}
\mathsf{NewRed}(M_{i+1}) &=& \{ R\in \mathsf{Redex}(M_{i+1}) 
\mid R\not\in \mathsf{Res}(\mathsf{Redex}(M_i)/R_i)\}.
\end{eqnarray*}
A redex occurrence $R_j\sqsubseteq M_j$ $(1\leq j)$ 
in $\sigma$ is called new 
if $R_j\in \mathsf{NewRed}(M_i)$ for some $i\leq j$. 
The reduction path $\sigma$ contains $k$ reductions of new redexes 
if $\sigma$ contracts $k$ of the new redexes.
\end{definition}

\section{New proof of the Church-Rosser theorem for $\beta$-equality}

\begin{proposition}[Complete development]\label{prop:CompleteDV}
We have $M\twoheadrightarrow^l M^*$
where $l\leq \frac{1}{2}\vert M\vert -1$ for $\vert M\vert \geq 4$.
\end{proposition}
{\it Proof.} By induction on the structure of $M$. 
Otherwise by the minimal complete development \cite{HS2008} 
with respect to $\mathsf{Redex}(M)$, 
where $l\leq\sharp \mathsf{Redex}(M) \leq \frac{1}{2}\vert M\vert-1$ 
for $\vert M\vert\leq 4$ by Lemma \ref{lemma:redex}.
\hfill $\Box$
\begin{definition}[Iteration of exponentials ${\mathbf 2}_n^m$, 
${\mathsf F}(m,n)$] Let $m$ and $n$ be natural numbers.
\begin{enumerate}
\item (1) ${\mathbf 2}_0^m = m$; ~  
 (2) ${\mathbf 2}_{n+1}^m = 2^{{\mathbf 2}_n^m}$.

\item (1) $\mathsf{F}(m, 0) = m$; ~ 
(2) $\mathsf{F}(m, n+1) = 2^{\mathsf{F}(m, n) -1}$.

\end{enumerate}
\end{definition}
\begin{proposition}[Length to $M^{n*}$]\label{prop:LengthToNStar}
If $M\twoheadrightarrow M^* \twoheadrightarrow \cdots 
\twoheadrightarrow M^{n*}$,
then the reduction length $l$ with $M\twoheadrightarrow^l M^{n*}$ 
is bounded by $\mathsf{Len}(\vert M\vert, n)$, 
such that 
\[
\mathsf{Len}(\vert M\vert, n) = \left\{
\begin{array}{rl}
0, &  \quad\mbox{for $n=0$}\\
 \displaystyle{\frac{1}{2} \sum_{k=0}^{n-1}
 \mathsf{F}(\vert M\vert, k) - n}, 
& \quad\mbox{for $n\geq 1$}
\end{array}\right. 
\] 
and then we have $\mathsf{Len}(\vert M\vert, n)< {\bf 2}_{n-1}^{\vert M\vert}$
for $n\geq 1$.  
\end{proposition}
{\it Proof.} 
From Lemma \ref{lemma:sizeofstar}, 
we have $\vert M^*\vert \leq 2^{\vert M\vert -1}$,
and hence 
$\vert M^{k*}\vert \leq \mathsf{F}(\vert M\vert, k) 
< {\bf 2}_{k}^{\vert M\vert}$ for $k\geq 1$.  
Let $M \twoheadrightarrow^{l_1} M^* \twoheadrightarrow^{l_2} \cdots 
\twoheadrightarrow^{l_n} M^{n*}$. 
Then from Proposition \ref{prop:CompleteDV}, 
each $l_k$ is bounded by ${\mathsf F}(\vert M\vert, k-1)$: 
$$
l_k ~~ \leq~~  \frac{1}{2}\vert M^{(k-1)*}\vert -1
 ~ \leq ~ \frac{1}{2}{\mathsf F}(\vert M\vert, k-1) -1
$$
Therefore, $l$ is bounded by $\mathsf{Len}(\vert M\vert, n)$
that is smaller than ${\bf 2}_{n-1}^{\vert M\vert}$ for $n\geq 1$.
$$
\hspace{15mm}
l ~~ \leq~~ \sum_{k=1}^n l_k 
 ~ \leq ~ 
\frac{1}{2}\sum_{k=0}^{n-1}\mathsf{F}(\vert M\vert, k) -n 
 ~ = ~ \mathsf{Len}(\vert M\vert, n)
 ~ < ~ 
\frac{1}{2}\sum_{k=0}^{n-1}{\mathbf 2}_k^{\vert M\vert} - n
 ~ < ~  {\mathbf 2}_{n-1}^{\vert M\vert} - n 
\hspace{18mm} \Box
$$
\begin{lemma}[(Weak) Cofinal property]\label{lem:WeakCR} 
If $M\to N$ then $N\twoheadrightarrow^l M^*$ 
where $l \leq \frac{1}{2}\vert N\vert -1$ for $\vert N\vert\geq 4$.
\end{lemma}
{\it Proof.}
By induction on the derivation of $M\to N$.
\hfill $\Box$
\begin{lemma}\label{lemma:StarSubst}
$M^*[x:= N^*] \twoheadrightarrow^l (M[x:= N])^*$
with $l\leq \vert M^*\vert-1$. 
\end{lemma}
{\it Proof.} 
By induction on the structure of $M$. 
We show one case $M$ of $M_1M_2$. 
\begin{enumerate}
\item Case $M_1 \equiv \lambda y.M_3$ for some $M_3$: 
\begin{eqnarray*}
((\lambda y.M_3)M_2)^* [x:= N^*]
 &=& M_3^*[x:= N^*] [y:= M_2^*[x:= N^*]]
\\
 &\twoheadrightarrow^{m_1} &
 M_3^*[x:= N^*] [y:= (M_2[x:= N])^*] 
 ~ \mbox{ by IH1}
\\
 &\twoheadrightarrow^{m_2} &
 (M_3[x:= N])^* [y:= (M_2[x:= N])^*] 
 ~ \mbox{ by IH2}
\end{eqnarray*}
Here, IH1 is $M_2^*[x:= N^*]\;\twoheadrightarrow^{n_1}\; (M_2[x:= N])^*$
with $n_1\;\leq\;\vert M_2^*\vert -1$, and 
then we have 
$m_1$ $=$ $\sharp(y\in(M_3^*[x:=N^*]))\times n_1$ 
from Lemma \ref{lemma:substitution}. 

IH2 is $M_3^*[x:= N^*] \twoheadrightarrow^{m_2} (M_3[x:= N])^*$
with $m_2 \leq \vert M_3^*\vert -1$. 
Hence, 
\begin{eqnarray*}
l &=& m_1 + m_2 
\\
&\leq& \sharp(y\in (M_3^*[x:=N^*]))\times (\vert M_2^*\vert -1) +\vert M_3^*\vert-1
\\
& = &
\sharp(y\in M_3^*)\times (\vert M_2^*\vert -1) + \vert M_3^*\vert -1
~ \mbox{ since $y\not\in\mathrm{FV}(N^*)$}
\\
&=& \vert M_3^*[y:= M_2^*]\vert -1. 
\end{eqnarray*}
\item Case $M_1\not\equiv \lambda y.M_3$:
\begin{enumerate}
\item Case $(M_1[x:=N]) \equiv (\lambda z.P)$ for some $P$:  
\begin{eqnarray*}
(M_1^*[x:= N^*])(M_2^*[x:= N^*])
 & \twoheadrightarrow^m & (M_1[x:= N])^* (M_2[x:= N])^* 
 ~ \mbox{ by IH}
\\
 & =  &
 (\lambda z. P^*) (M_2[x:= N])^* 
\\
 & \twoheadrightarrow^1 &
  P^* [z:= (M_2[x:= N])^*] 
\\
 &=& ((M_1 M_2)[x:=N])^*
\end{eqnarray*}
Now, IH are 
$M_1^*[x:= N^*] \twoheadrightarrow^{n_1} (M_1[x:= N])^*$
with $n_1 \leq \vert M_1^*\vert -1$, and 
$M_2^*[x:= N^*]$ $\twoheadrightarrow^{n_2}$ $(M_2[x:= N])^*$
with $n_2 \leq \vert M_2^*\vert -1$. Hence,  
\begin{eqnarray*}
l & = & m +1 
\\
& \leq & \vert M_1^*\vert -1 + \vert M_2^*\vert -1 +1 
\\
&<& \vert M_1^* M_2^*\vert -1. 
\end{eqnarray*}

\item Case $(M_1[x:=N]) \not\equiv (\lambda z.P)$: 

This case is handled similarly to the above case, and then
\begin{eqnarray*}
l &\leq& m 
\\
&=& \vert M_1^*\vert -1 + \vert M_2^*\vert -1 
\\
&<& \vert M_1^* M_2^*\vert -1. 
 \hspace{95mm} \Box
\end{eqnarray*}
\end{enumerate}
\end{enumerate}
\begin{proposition}[Monotonicity]\label{prop:Mono}
If $M\to N$ then 
$M^*\twoheadrightarrow^l N^*$
with $l\leq \vert M^*\vert-1$.
\end{proposition}
{\it Proof.} 
By induction on the derivation of $M\to N$. 
We show some of the interesting cases.
\begin{enumerate}
\item Case of $(\lambda x.M)N \to M[x:=N]$: 
\begin{eqnarray*}
((\lambda x.M)N)^* 
 &=& M^*[x:= N^*]  
\\
& \twoheadrightarrow^m & (M[x:= N])^*
\end{eqnarray*}
From Lemma \ref{lemma:StarSubst},   
we have $m\leq \vert M^*[x:= N^*]\vert -1 = \vert ((\lambda x.M)N)^*\vert -1$.

\item Case of $PM \to PN$ from $M\to N$: 
\begin{enumerate}
\item Case of $P\equiv\lambda x.P_1$ for some $P_1$:
\begin{eqnarray*}
((\lambda x.P_1)M)^* 
 &=& P_1^*[x:= M^*] 
\\
&\twoheadrightarrow^m& P_1^*[x:= N^*] 
~ \mbox{ by IH}
\\
 &=&((\lambda x.P_1)N)^* 
\end{eqnarray*}
%
Here, IH is $M^* \twoheadrightarrow^n N^*$ with $n\leq \vert M^*\vert-1$, 
and $m = \sharp(x\in P_1^*)\times n$ from Lemma \ref{lemma:substitution}. 
Hence, 
\begin{eqnarray*}
l  &=& m 
\\
&\leq& \sharp(x\in P_1^*)\times (\vert M^*\vert -1)
\\
&\leq&
\vert P_1^*\vert + \sharp(x\in P_1^*)\times (\vert M^*\vert -1) -1
\\
&=& \vert P_1^*[x:= M^*] \vert -1. 
\end{eqnarray*}
\item Case of $P\not\equiv\lambda x.P_1$:
Similarly handled. 
\hfill $\Box$
\end{enumerate}
\end{enumerate}
\begin{lemma}[Main lemma]\label{th:main}
Let $M =_\beta^k N$ with length $k=l+r$, where 
$r$ is the number of occurrences of right arrow $\to$ 
in $=_\beta^k$, and 
$l$ is that of left arrow $\leftarrow$ 
in $=_\beta^k$. 
Then we have both $M^{r*}\twoheadleftarrow N$
and $M \twoheadrightarrow N^{l*}$.
\end{lemma}
{\it Proof.} By induction on the length of $=_\beta^k$. 
\begin{enumerate}
\item[(1)] Case of $k=1$ is handled by Lemma \ref{lem:WeakCR}.

\item[(2-1)] Case of $(k+1)$, where $M =_\beta^k M_k \to M_{k+1}$: 

From the induction hypothesis, 
we have $M_k \twoheadrightarrow M^{r*}$ 
and $M \twoheadrightarrow M_k^{l*}$ 
where $l+r = k$. 

From $M_k \to M_{k+1}$, Lemma \ref{lem:WeakCR} gives 
$M_{k+1}\twoheadrightarrow M_k^*$, and 
then $M_k^{*} \twoheadrightarrow M^{(r+1)*}$ from 
the induction hypothesis $M_k \twoheadrightarrow M^{r*}$ 
and Proposition \ref{prop:Mono}.
Hence, we have $M_{k+1} \twoheadrightarrow M^{(r+1)*}$. 
On the other hand, 
we have $M_k^{l*} \twoheadrightarrow M_{k+1}^{l*}$ 
from $M_k\to M_{k+1}$ and the repeated application of 
Proposition \ref{prop:Mono}. 
Then the induction hypothesis $M\twoheadrightarrow M_k^{l*}$ 
derives $M\twoheadrightarrow M_{k+1}^{l*}$, 
where $l + (r+1) = k+1$. 

\item[(2-2)] Case of $(k+1)$, where $M=_\beta^k M_k \leftarrow M_{k+1}$: 

From the induction hypothesis, 
we have $M_k \twoheadrightarrow M^{r*}$ 
and $M \twoheadrightarrow M_k^{l*}$ 
where $l+r = k$, and hence $M_{k+1}\twoheadrightarrow M^{r*}$. 
From $M_{k+1}\to M_k$ and Lemma \ref{lem:WeakCR}, 
we have $M_k\twoheadrightarrow M_{k+1}^*$, 
and then $M_k^{l*} \twoheadrightarrow M_{k+1}^{(l+1)*}$.
Hence, $M\twoheadrightarrow M_{k+1}^{(l+1)*}$ from 
the induction hypothesis $M \twoheadrightarrow M_k^{l*}$, 
where $(l+1) + r = k+1$.  
\hfill $\Box$
\end{enumerate}
Given $M_0 =_\beta^k M_k$ with reduction sequence $[M_0, \ldots, M_k]$,
then for natural numbers $i$ and $j$ with $0\leq i\leq j\leq k$, 
we write $\sharp r[i, j]$ for the number of occurrences of right arrow 
$\to$ which appears in $M_i =_\beta^{(j-i)} M_j$, and $\sharp l[i, j]$ for that 
of left arrow $\leftarrow$ in $M_i =_\beta^{(j-i)} M_j$. 
In particular, we have $\sharp l[0,k] + \sharp r[0,k] = k$. 
\begin{corollary}[Main lemma refined]\label{cor:RefinedTheorem}
Let $M_0 =_\beta^k M_k$ with reduction sequence  $[M_0, M_1, \ldots, M_k]$.
Let $r = \sharp r[0, k]$ and $l = \sharp l[0,k]$. 
Then we have $M_0\twoheadrightarrow M_{r}^{m_l *}$ 
and $M_{r}^{m_l *} \twoheadleftarrow M_k$, where  
$m_l = \sharp l[0, r] \leq \min\{l, r\}$. 
\end{corollary}
{\it Proof.} 
From the main lemma, we have two reduction paths such that 
$M_0\twoheadrightarrow M_k^{l*}$ 
and $M_0^{r*} \twoheadleftarrow M_k$, 
where the paths have a crossed point that is the term $M_{r}^{n*}$ 
for some $n\leq k$ as follows:  
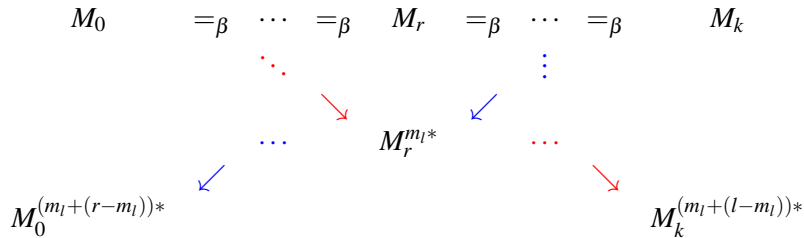
\[
\begin{array}{ccccccccc}
M_0 &  =_\beta &  \cdots & =_\beta & M_{r}  
 &  =_\beta &\cdots &=_\beta & M_k
\\
 & & \textcolor{red}{\ddots} & & &  & \textcolor{blue}{\vdots} & & 
\\
   & & &  \textcolor{red}{\searrow} & & \textcolor{blue}{\swarrow} &  & & 
\\
 &  & \textcolor{blue}{\cdots} & & M_{r}^{m_l *} &  & {\color{red}\cdots} & & 
\\
 & \textcolor{blue}{\swarrow} &  &  &  
 & & &  \textcolor{red}{\searrow}  & 
\\
M_0^{(m_l + (r-m_l))*}   &  &   &   & &  & & & M_k^{(m_l + (l-m_l))*}
\end{array}
\]
Let \textcolor{black}{$m_l$} be $\sharp l[0, r]$,
then $\sharp l[r, k] = (l-m_l)$ and  $\sharp r[r, k] = m_l$. 
Hence, from the main lemma, 
we have $M_0 \twoheadrightarrow M_r^{m_l*} \twoheadleftarrow M_k$ 
where $m_l \leq \min\{l, r\}$.  
Moreover, 
we have $M_r \twoheadrightarrow M_{k}^{(l-m_l)*}$ by the main lemma again,  
and then $M_r^{m_l*} \twoheadrightarrow M_{k}^{((l-m_l) + m_l)*}$ 
from the repeated application of Proposition \ref{prop:Mono}. 
Therefore, we indeed have $M_0 \twoheadrightarrow M_r^{m_l*} 
\twoheadrightarrow M_{k}^{l*}$. 
Similarly, we have $M_0^{r*} \twoheadleftarrow M_r^{m_l*} \twoheadleftarrow M_k$
as well. \hfill $\Box$
\begin{example} 
We demonstrate a simple example of $M_0 =_\beta^4 M_4$ with length $4$, 
and list $2^4$ patterns of the reduction graph 
consisting of the sequence $[M_0, M_1, M_2, M_3, M_4]$.
The sixteen patterns can be classified into 5 groups, in which 
$M_0$ and $M_4$ have a pair of the same common reducts  
$\langle M_0^{r*}, M_4^{l*}\rangle$ where $r+l = 4$: 
\begin{enumerate}
\item Common reducts $\langle M_0^{4*}, M_4^{0*}\rangle$ 
and a crossed point $M_4^{m_l*} \equiv M_4^{0*}$:

(1) $M_0 \to M_1 \to M_2\to M_3 \to M_4$. 

\item Common reducts $\langle M_0^{3*}, M_4^{*}\rangle$ and 
crossed points $M_3^{m_l*}$ of two kinds: 

(1) $M_0 \leftarrow M_1 \to M_2\to M_3\to M_4$;  
 ~ ~ (2) $M_0 \to M_1 \leftarrow M_2\to M_3 \to M_4$ with $M_3^{m_l*}\equiv M_3^*$;  

(3) $M_0 \to M_1 \to M_2\leftarrow M_3 \to M_4$;   
 ~ ~ (4) $M_0 \to M_1 \to M_2\to M_3 \leftarrow M_4$ with $M_3^{m_l*}\equiv M_3^{0*}$.

\item $\langle M_0^{2*}, M_4^{2*}\rangle$ and 
crossed points $M_2^{m_l*}$ of three kinds:

(1) $M_0 \leftarrow M_1 \to M_2\leftarrow M_3 \to M_4$; ~ ~  
(2) $M_0 \leftarrow M_1 \leftarrow M_2\to M_3 \to M_4$ 
with $M_2^{m_l*}\equiv M_2^{2*}$; 

(3) $M_0 \leftarrow M_1 \to M_2\to M_3 \leftarrow M_4$; ~ ~ 
(4) $M_0 \to M_1 \leftarrow M_2\to M_3 \leftarrow M_4$
 with $M_2^{m_l*}\equiv M_2^{*}$;

(5) $M_0 \to M_1 \leftarrow M_2\leftarrow M_3 \to M_4$; ~ ~  
(6) $M_0 \to M_1 \to M_2\leftarrow M_3 \leftarrow M_4$ 
  with $M_2^{m_l*}\equiv M_2^{0*}$.

\item $\langle M_0^{*}, M_4^{3*}\rangle$ and 
crossed points $M_1^{m_l*}$ of two kinds:

(1) $M_0 \leftarrow M_1 \to M_2 \leftarrow M_3 \leftarrow M_4$; ~ ~  
(2) $M_0 \leftarrow M_1 \leftarrow M_2 \leftarrow M_3 \to M_4$ 
 with $M_1^{m_l*}\equiv M_1^{*}$; ~ ~   

(3) $M_0 \leftarrow M_1 \leftarrow M_2 \to M_3 \leftarrow M_4$;  ~ ~ 
(4) $M_0 \to M_1 \leftarrow M_2\leftarrow M_3 \leftarrow M_4$
 with $M_1^{m_l*}\equiv M_1^{0*}$.  

\item $\langle M_0^{0*}, M_4^{4*}\rangle$ and 
a crossed point $M_0^{m_l*}\equiv M_0^{0*}$: 

(1) $M_0 \leftarrow M_1 \leftarrow M_2 \leftarrow M_3 \leftarrow M_4$. 

\end{enumerate}
Observe that a crossed point $M_r^{m_l*}$ in 
Corollary \ref{cor:RefinedTheorem} gives 
a ``good'' common contractum such that the number $m_l$, i.e., iteration of 
the translation $*$ is minimum, see also the trivial cases above; 
Case 1, Case 2 (4), Case 3 (6), Case 4 (4), and Case 5. 
Consider two reduction paths:  
(i) a reduction path from $M_r^{m_l*}$ to $M_0^{r*}$, 
and (ii) a reduction path from $M_r^{m_l*}$ to $M_k^{l*}$, 
see the picture in the proof of Corollary \ref{cor:RefinedTheorem}.
In general, 
the reduction paths (i) and (ii) form 
the boundary line between common contractums and non-common ones.  
Let $B$ be a term in the boundary (i) or (ii). Then 
any term $M$ such that $B \twoheadrightarrow M$ is a common contractum 
of $M_0$ and $M_k$. In this sense, the term $M_r^{m_l*}$ 
where $0\leq m_l \leq \min\{l, r\}$ 
can be considered as an optimum common reduct of $M_0$ and $M_k$ 
in terms of Takahashi translation.  
Moreover, the refined lemma gives a divide and conquer method
such that $M_0=_\beta^k M_k$ is divided into $M_0 =_\beta^r M_r$
and $M_r =_\beta^l M_k$, where the base case is a valley such that   
$M_0 \twoheadrightarrow M_r \twoheadleftarrow M_k$ 
with minimal $M_r$ and $m_l=0$, as shown by the trivial cases above.
\end{example}
The results of Lemma \ref{th:main} 
and Corollary \ref{cor:RefinedTheorem} 
can be unified as follows. 
The main theorem shows that every term in the reduction sequence $ls$ of 
$M_0=_\beta^k M_k$ generates a common contractum:
For every term $M $ in $ls$, there exists a natural number $n\leq \max\{l, r\}$ 
such that $M^{n*}$ is a common contractum of $M_0$ and $M_k$. 
Moreover, 
there exist a term $N$ in $ls$ and a natural number $m\leq \min\{l,r\}$ 
such that $N^{m*}$ 
is a common contractum of all the terms in $ls$.
\begin{theorem}[Main theorem for $\beta$-equality]\label{th:Church-Rosser}
Let $M_0 =_\beta^k M_{k}$ with reduction sequence $[M_0, \ldots, M_k]$.
Let $l = \sharp l[0,k]$ and $r =\sharp r[0,k]$. 
Then there exist the following common reducts: 
\begin{enumerate}
\item We have 
$M_0\twoheadrightarrow M_{r-i}^{\sharp r[r-i,k] *}$
and $M_{r-i}^{\sharp r[r-i,k] *} \twoheadleftarrow M_k$
for each $i=0, \ldots, r$. 
We also have $M_0\twoheadrightarrow M_{r+j}^{\sharp l[0,r+j] *}$
and $M_{r+j}^{\sharp l[0,r+j] *} \twoheadleftarrow M_k$
for each $j=0, \ldots, l$.

\item For every term $M$ in the reduction sequence, we have 
$M\twoheadrightarrow M_r^{m_l*}$ where $m_l = \sharp l[0,r]$. 
\end{enumerate}
\end{theorem}
{\it Proof.} Both 1 and 2 are proved similarly 
from Lemma \ref{th:main}, Corollary \ref{cor:RefinedTheorem}, 
and monotonicity. We show the case 2 here.  
Let $M_i$ be a term in the reduction sequence of $M_0 =_\beta^k M_k$ 
where $0\leq i\leq r$.  
Take $a = \sharp r[0, i]$, then $M_a^{\sharp l[0,a]}$ is 
a crossed point of $M_0\twoheadrightarrow M_i^{\sharp l[0,i]*}$ 
and $M_i \twoheadrightarrow M_0^{\sharp r[0,i]*}$. 
From $M_i \twoheadrightarrow M_r^{\sharp l[i, r]*}$ 
and monotonicity, 
we have $M_i^{\sharp l[0,i]*} \twoheadrightarrow M_r^{m_l *}$
where $m_l = \sharp l[0, i] + \sharp l[i, r]$. 
Hence, we have $M_i \twoheadrightarrow 
M_a^{\sharp l[0,a]*} \twoheadrightarrow 
M_i^{\sharp l[0,i]*} \twoheadrightarrow M_r^{m_l *}$. 
The case of $r\leq i\leq k$ is also verified similarly. 
 \hfill $\Box$
\\
Note that the case of $i=r$ and $j=l$ implies the main lemma, 
since $\sharp r[0,k] = r$ and 
$\sharp l[0,r+l] = \sharp l[0,k]= l$. 
Note also that the case of $i=0 =j$ implies the refinement, 
since $\sharp l[0, r] = m_l = \sharp r[r,k]$.
\begin{corollary}[Church-Rosser theorem for $\beta$-reduction]
\label{th:CR-reduction} 
Let $P_n\leftarrow \cdots \leftarrow P_1 \leftarrow M
\to Q_1\to\cdots \to Q_m$  $(1\leq n\leq m)$.  
Then we have 
$P_n \twoheadrightarrow Q_{m}^{n*}$ 
and $Q_{m}\twoheadrightarrow Q_{m}^{n*}$. 
We also have $P_n \twoheadrightarrow Q_{(m-n)}^{n*}$ 
and $Q_{m}\twoheadrightarrow Q_{(m-n)}^{n*}$. 
\end{corollary}
{\it Proof.} From the main lemma and the refinement
where $Q_0\equiv M$. \hfill $\Box$
\begin{theorem}[Improved Church-Rosser theorem for $\beta$-reduction]\label{th:ImprovedCR}
Let $P_n\leftarrow \cdots \leftarrow P_1 \leftarrow M
\to Q_1\to\cdots \to Q_{m}$ $(1\leq n\leq m)$.   
If 
$P_n\leftarrow \cdots \leftarrow P_1 \leftarrow M$ contains 
$a$-times reductions of new redexes $(0\leq a \leq n-1)$,  
and $M\to Q_1\to\cdots \to Q_{m}$
contains $b$-times reductions of new redexes $(0\leq b \leq m-1)$, 
then we have both 
$P_n \twoheadrightarrow Q_{m}^{(a+1)*}$ 
and $Q_{m}\twoheadrightarrow P_{n}^{(b+1)*}$. 
\end{theorem}
{\it Proof.} 
We show the claim that 
if a reduction path $\sigma$ of 
$R_0 R_1\ldots R_n: M\equiv M_0\to M_1 \to \cdots \to M_{n+1}$
contains $a$-times reductions of new redexes $(1\leq a\leq n-1)$
then $M_{n+1} \twoheadrightarrow M^{(a+1)*}$, 
from which the theorem is derived by repeated application of 
Proposition \ref{prop:Mono}.

We prove the claim by induction on $a$. 
\begin{enumerate}
\item Case of $a=0$:

We have 
$R_0 R_1\ldots R_n: M\equiv M_0\to  M_1 \to \cdots \to M_{n+1}$,  
where none of $R_i$ $(0\leq i\leq n)$ is a new redex. 
The reduction path is a development of $M$ with respect to a subset of 
$\mathsf{Redex}(M)$. 
Then we have $M_{j} \twoheadrightarrow M^*$ $(0\leq j\leq n+1)$,  
since all developments of $\mathsf{Redex}(M)$ are finite 
\cite{Hindley1978,Barendregt1984} 
and end with some $N$ such that $N\twoheadrightarrow M^*$.
 
\item Case of $a=k+1$: 

We have 
$R_0R_1\ldots R_{n-1}R_nR_{n+1}\ldots R_m 
: M\equiv M_0\to  M_1 \to \cdots \to M_n\to
M_{n+1} \to \cdots \to M_{m+1}$ $(m\geq 0)$,
where $R_0R_1\ldots R_{n-1}: M\equiv M_0\to  M_1 \to \cdots \to M_n$ contains 
$k$ reductions of new redexes $(0\leq k\leq n-1)$.  
Moreover, the redex ${R_n}$ is a new redex, and   
$R_{n+1} \ldots R_m : M_{n+1} \to \cdots \to M_{m+1}$ contains no new redexes.
Then the reduction path 
$R_nR_{n+1} \ldots R_m : M_n\to M_{n+1} \to \cdots \to M_{m+1}$ 
is a development of $M_{n}$ with respect to a subset of 
$\mathsf{Redex}(M_{n})$, and hence $M_{m+1} \twoheadrightarrow M_n^*$.
On the other hand, from the induction hypothesis applied to 
the reduction path 
$R_0 R_1 \ldots R_{n-1} : M\equiv M_0\to M_1 \to \cdots \to M_n$ 
with $k$ reductions of new redexes,
we have $M_n \twoheadrightarrow M^{(k+1)*}$. 
Therefore, we have $M_{m+1} \twoheadrightarrow M^{(k+2)*}$ 
by repeated application of Proposition \ref{prop:Mono}.
\hfill $\Box$
\end{enumerate}

\section{Quantitative analysis and comparison with related results}

\subsection{Measure functions}

For quantitative analysis, we list important measure functions,
$\mathsf{TermSize}$, $\mathsf{Mon}$, and $\mathsf{Rev}$. 
\begin{definition}[TermSize]\label{def:TermSize}
We define $\mathsf{TermSize}(M =_\beta N)$ 
by induction on the derivation. 
\begin{enumerate}
\item If $M \twoheadrightarrow^r N$ then 
$\mathsf{TermSize}(M =_\beta N) = 8(\frac{\vert M\vert}{8})^{2^r}$. 
\item If $M =_\beta N$ is derived from $N =_\beta M$,
then define $\mathsf{TermSize}(M =_\beta N)$ by 
$\mathsf{TermSize}(N =_\beta M)$.
\item If $M =_\beta N$ is derived from $M =_\beta P$ and $P =_\beta N$,
then define $\mathsf{TermSize}(M =_\beta N)$ as follows: 
$\max\{ \mathsf{TermSize}(M =_\beta P),  \mathsf{TermSize}(P =_\beta N)\}$.
\end{enumerate}
\end{definition}

\begin{proposition}[TermSize]\label{prop:TermSize}
Let $M_0 =_\beta^k M_k$ with reduction sequence $ls$. 
Then $\vert M \vert \leq \mathsf{TermSize}(M_0 =_\beta^k M_k)$ 
for each term $M$ in $ls$, 
and $\mathsf{TermSize}(M_0 =_\beta^k M_k) \leq \vert N\vert^{2^k}$
for some term $N$ in $ls$.
\end{proposition}
{\it Proof.} By induction on the derivation of $=_\beta$ 
together with Definition \ref{def:TermSize} and Proposition \ref{prop:length}. 
\hfill $\Box$
\begin{definition}[Monotonicity]\label{def:n-mono}
\[
\mathsf{Mon}(\vert M\vert , m, n) = \left\{
\begin{array}{ll} 
\displaystyle{2^{\vert M\vert^{2^m}}}, &  \quad\mbox{for $n=1$}
\\
 \displaystyle{ 2 ^{2^{[ 2^{\mathsf{Mon}(\vert M\vert, m, n-1)} 
\times \mathbf{2}_{(n-2)}^{\vert M\vert}]}}}, 
& \quad\mbox{for $n > 1$}
\end{array}\right. 
\] 
\end{definition}
\begin{proposition}[Monotonicity]\label{prop:n-mono}
If $M\twoheadrightarrow^{m} N$,
then $M^{n*} \twoheadrightarrow^{l} N^{n*}$ 
with $l \leq \mathsf{Mon}(\vert M\vert, m, n)$.
\end{proposition}
{\it Proof.} 
By induction on $n$. 
\begin{enumerate}
\item Case of $n = 1$: 

If $M \twoheadrightarrow^m M_m$, then 
$M^* \twoheadrightarrow^l M_m^*$ with 
$\displaystyle{l \leq 2^{\vert M\vert^{2^m}}}$. 
Indeed, from Proposition \ref{prop:length}, we have 
$\vert M_m\vert < \vert M\vert^{2^m}$.  
If $M_0 \to M_1$ then we have $M_0^* \twoheadrightarrow^{l_1} M_1^*$
with $l_1 < 2^{\vert M_0\vert}$ 
from Proposition \ref{prop:Mono} and Lemma \ref{lemma:sizeofstar}.
Hence, from $M_0 \to M_1\to \cdots \to M_m$, 
we have $M_0^* \twoheadrightarrow^{l_1} 
M_1^* \twoheadrightarrow^{l_2} \cdots \twoheadrightarrow^{l_m} M_m^*$ where 
$$
l ~~ = ~~  \sum_{i=1}^{m} l_i ~~ < ~~ \sum_{i=0}^{m-1} 2^{\vert M_i\vert}
 ~~ < ~~ \sum_{i=0}^{m-1} 2^{\vert M_0\vert^{2^{i}}} 
~~ < ~~ 2^{\vert M_0\vert^{2^{m}}}.
$$ 
\item Case of $n\geq 1$: 

From the induction hypothesis, we have 
$M^{n*} \twoheadrightarrow^{l} N^{n*}$ 
with $l < \mathsf{Mon}(\vert M\vert, m, n)$. 
Therefore, we have 
$M^{(n+1)*} \twoheadrightarrow^{l'} N^{(n+1)*}$ 
with 
$$
 \hspace{30mm}
l'  ~~<~~ 2^{\vert M^{n*}\vert^{2^l}}
 ~~ < ~~ 
2^{\vert M^{n*}\vert^{2^{\mathsf{Mon}(\vert M\vert, m, n)}}},
~~ \mbox{ where $\vert M^{n*}\vert < {\mathbf 2}_n^{\vert M\vert}$}. 
 \hspace{30mm} \Box
$$
\end{enumerate}
\begin{lemma}[Cofinal property]\label{lem:ExtWeakCR}
If $M\twoheadrightarrow^n N$ $(n\geq 1)$, 
then $N\twoheadrightarrow^l M^{n*}$ with 
$l < \mathsf{Rev}(\vert M\vert, n)$ as follows: 
\[
\mathsf{Rev}(\vert M\vert, n) = \left\{ 
\begin{array}{ll}
 \frac{1}{2}\vert M\vert^2, &\quad\mbox{for $n=1$}
\\
\frac{1}{2}\vert M\vert^{2^n} 
 + 2^{\vert M\vert^{2^{[n-1 + \mathsf{Rev}(\vert M\vert, n-1)]}}}, 
&\quad\mbox{for $n > 1$}
\end{array}
\right.
\]
\end{lemma}
%
%
{\it Proof.} 
The case $\mathsf{Rev}(\vert M\vert, 1)$ is by Lemma \ref{lem:WeakCR}.  
For $n>1$, $\mathsf{Rev}(\vert M\vert, n)$ follows 
$\mathsf{Mon}(\vert M\vert, n, 1)$ from Proposition \ref{prop:n-mono} 
and $\vert N\vert < \vert M\vert^{2^{n}}$ from 
 Proposition \ref{prop:length}.
\hfill $\Box$
%
%

\subsection{Quantitative analysis of Church-Rosser for $\beta$-reduction}

We show two bound functions $f(l, \vert M\vert, r)= \langle m, n\rangle$ 
such that 
for the peak $N_1\twoheadleftarrow^l M\twoheadrightarrow^r N_2$, 
the valley size of 
$N_1\twoheadrightarrow^a P\twoheadleftarrow^b N_2$ for some $P$ 
is bounded by $a \leq m$ and $b\leq n$.
The first function $\mathsf{CR\mbox{-}red}(l, M, r) 
= \langle m, N_1^{r*}, n\rangle$ 
provides a common reduct $N_1^{r*}$,  
following the proof of the main lemma with $\mathsf{Mon}$.
The second one 
$\mathsf{V\mbox{-}size}(l, M, r) = \langle m, M^{r*}, n\rangle$
gives a common reduct $M^{r*}$ simply using $\mathsf{Rev}$
provided that $l\leq r$.
\begin{definition}[$\mathsf{CR\mbox{-}red}$]
\begin{enumerate}
\item $\mathsf{CR\mbox{-}red}(l, M, 1) 
= \langle \frac{1}{2}\vert M\vert^{2^l}, N_1^*, \frac{1}{2}\vert M\vert^2 +
2^{\vert M\vert^{2^l}}\rangle$

\item $\mathsf{CR\mbox{-}red}(l, M, r) = $

{\tt let}~  
$\langle m, N_1^{(r-1)*}, n\rangle$ ~ {\tt be} ~  
$\mathsf{CR\mbox{-}red}(l, M, r-1)$ ~ 
{\tt in} ~  
$\langle \mathbf{2}_{(r-1)}^{\vert M\vert^{2^l}}, N_1^{r*}, 
\frac{1}{2}\vert M\vert^{2^r} + 2^{\vert M\vert^{2^{[r -1 + n]}}}\rangle$
 for $r>1$

\end{enumerate}
\end{definition}

\begin{proposition}[$\mathsf{CR\mbox{-}red}$]\label{prop:CR-red}
If $N_1\twoheadleftarrow^l M\twoheadrightarrow^r N_2$, 
then we have $\mathsf{CR\mbox{-}red}(l, M, r) = \langle m, N_1^{r*}, n\rangle$
such that \\
$N_1\twoheadrightarrow^a N_1^{r*} \twoheadleftarrow^b N_2$
with $a \leq m$ and $b\leq n$. 
\end{proposition}
{\it Proof.} By induction on $r$. 
\begin{enumerate}
\item Case $r=1$: 

We have $M^*\twoheadleftarrow^a N_2$ with $a\leq\frac{1}{2}\vert N_2\vert
\leq\frac{1}{2}\vert M\vert^2$. 
Then $N_1^*\twoheadleftarrow^b M^*$ with 
$b\leq \mathsf{Mon}(\vert M\vert, l, 1) = 2^{\vert M\vert^{2^l}}$. 
On the other hand, we have a common contractum $N_1^*$ such that 
$N_1\twoheadrightarrow^c N_1^*$ with 
$c\leq \frac{1}{2}\vert N_1\vert \leq \frac{1}{2}\vert M\vert^{2^l}$.

\item Case of $r>1$: 

From the induction hypothesis, we have 
$\langle m, N_1^{(r-1)}, n\rangle = \mathsf{CR\mbox{-}red}(l, M, r-1)$
such that 
\\
$M\twoheadrightarrow^{(r-1)} N_3 \rightarrow N_2$ 
and $N_1^{(r-1)*}\twoheadleftarrow^b N_3$ with $b\leq n$ for some $N_3$. 
Then we have $N_3^*\twoheadleftarrow^c N_2$ with 
$c\leq \frac{1}{2}\vert N_2\vert \leq \frac{1}{2}\vert M\vert^{2^{r}}$,
and hence $N_1^{r*}\twoheadleftarrow^d N_3^*$ where 
$$
d ~~ \leq~~ \mathsf{Mon}(\vert N_3\vert, n, 1)
 ~~ \leq ~~  \mathsf{Mon}(\vert M\vert^{2^{(r-1)}}, n, 1)
 ~~ = ~~ 2^{(\vert M\vert^{2^{(r-1)}})^{2^n}} 
 ~~ = ~~ 2^{\vert M\vert^{2^{[r + n -1]}}}. 
$$
Therefore, we have a common reduct $N_1^{r*}$ such that  
$N_1\twoheadrightarrow^e N_1^{r*}$ 
with $e\leq \mathsf{Len}(\vert N_1\vert, r) 
\leq \mathbf{2}_{(r-1)}^{\vert M\vert^{2^l}}$.
\hfill $\Box$
\end{enumerate}
\begin{definition}[$\mathsf{V\mbox{-}size}$]
~ $\mathsf{V\mbox{-}size}(l, M, r) 
= \langle \mathsf{Rev}(\vert M\vert, l) + \mathbf{2}_{r-1}^{\vert M\vert}, 
M^{r*},  \mathsf{Rev}(M, r)\rangle$ for $1\leq l\leq r$.
\end{definition}
\begin{proposition}[$\mathsf{V\mbox{-}size}$]\label{prop:V-size}
If $N_1\twoheadleftarrow^l M\twoheadrightarrow^r N_2$ with $l\leq r$, 
then we have $\mathsf{V\mbox{-}size}(l, M, r) = \langle m, M^{r*}, n\rangle$
such that 
$N_1\twoheadrightarrow^a M^{r*} \twoheadleftarrow^b N_2$
with $a \leq m$ and $b\leq n$. 
\end{proposition}
{\it Proof.} Suppose that $l\leq r$. 
We have $N_1\twoheadrightarrow^a M^{l*}$ with 
$a\leq \mathsf{Rev}(\vert M\vert, l)$ and 
$M^{r*}\twoheadleftarrow^b N_2$ with 
$b\leq \mathsf{Rev}(\vert M\vert, r)$, respectively. 
From $l\leq r$, we have $M^{l*} \twoheadrightarrow^c M^{r*}$ 
where 
$$
\hspace{30mm} 
c ~~\leq ~~ \mathsf{Len}(\vert M^{l*}\vert, r-l)
  ~~ \leq ~~ \mathbf{2}_{r-l-1}^{\vert M^{l*}\vert}
  ~~ \leq ~~ \mathbf{2}_{r-l-1}^{\mathbf{2}_l^{\vert M\vert}}
 ~~ = ~~ \mathbf{2}_{r-1}^{\vert M\vert}.
\hspace{35mm} \Box
$$

On the other hand, 
Ketema and Simonsen \cite{KS2013} showed that 
an upper bound on the size of confluence diagrams in $\lambda$-calculus
is $\mathsf{bl}(l, \vert M\vert, r)$ for 
$P \twoheadleftarrow^l  M\twoheadrightarrow^r Q$. 
The valley size $a$ and $b$ of 
$P\twoheadrightarrow^{a} N \twoheadleftarrow^{b} Q$ 
for some $N$ is bounded by $\mathsf{bl}(l, \vert M\vert, r)$ as follows:
\[
\mathsf{bl}(l, \vert M\vert, r) = \left\{ 
\begin{array}{ll}
 \vert M\vert^{2^{[2^{l} + l + 2]}}, &\quad\mbox{for $r=1$}
\\
\vert M\vert^{2^{[2^{\mathsf{bl}(l, \vert M\vert, r-1) } +
\mathsf{bl}(l, \vert M\vert, r-1) 
+ r+1]}}, 
&\quad\mbox{for $r > 1$}
\end{array}
\right.
\]
Their proof method is based on the use of the so-called Strip Lemma, 
and in this sense our first method $\mathsf{CR\mbox{-}red}$ is rather 
similar to theirs. 
However, for a large term $M$, $\mathsf{bl}$ can give a shorter reduction 
length than that by $\mathsf{CR\mbox{-}red}$ 
from the shape of the functions. 
The reason can be expounded as follows: From given terms, we explicitly 
constructed a common reduct via $*$-translation, so that more redexes 
than a set of residuals can be reduced, compared with those of $\mathsf{bl}$. 
To overcome this point, an improved version of Theorem \ref{th:ImprovedCR} 
is introduced such that $*$-translation is applied 
only when new redexes are indeed reduced.

The basic idea of the second method $\mathsf{V\mbox{-}size}$ is essentially 
the same as the proof given in \cite{KMY2014}. 
In summary, the functions $\mathsf{bl}$ and $\mathsf{CR\mbox{-}red}$ 
including a common reduct are respectively 
defined by induction on the length of one side of the peak, 
and $\mathsf{V\mbox{-}size}$ is by induction on that of 
both sides of the peak. 
All the functions belong to 
the fourth level of the Grzegorczyk hierarchy.    

\subsection{Quantitative analysis of Church-Rosser for $\beta$-equality}

Let $M_0 =_\beta^k M_k$ with length $k = l + r$
where $l=\sharp l[0,k]$ and $r=\sharp r[0,k]$, and 
$\mathsf{M_{}}$ be $\mathsf{TermSize}( M_0 =_\beta^k M_k)$. 
Then we show a bound function $\mathsf{CR\mbox{-}eq}(M_0=_\beta^k M_k)
= \langle m, M_0^{r*}, n\rangle$ such that 
$M_0\twoheadrightarrow^a M_0^{r*}$ and $M_0^{r*}\twoheadleftarrow^b M_k$ 
with $a\leq m$ and $b\leq n$.
This analysis reveals the size of the valley described in Lemma \ref{th:main}. 
\begin{definition}\label{def:CRE}
Given $M_0 =_\beta^k M_k$ with length $k = l + r$  
where $l = \sharp l[0,k]$ and $r = \sharp r[0,k]$.
Let $\mathsf{M_{}}$ be $\mathsf{TermSize}( M_0 =_\beta^k M_k)$. 
A measure function $\mathsf{CR\mbox{-}eq}$ is defined by induction on 
the length of $=_\beta^k$, where $\cdot$ denotes an arbitrary term.
\begin{enumerate}
\item 
$\mathsf{CR\mbox{-}eq}(M_0 \leftarrow \cdot) = \langle 0, M_0^{0*}, 1\rangle$; 
 ~ 
$\mathsf{CR\mbox{-}eq}(M_0\rightarrow \cdot) 
= \langle \frac{1}{2}\vert M_0\vert, M_0^{*}, \frac{1}{2}\vert M_0\vert^2\rangle$

%
%
\item $\mathsf{CR\mbox{-}eq}(M_0=_\beta^{k} \cdot \leftarrow \cdot)  = $ 
${\tt let} ~ \langle a, M_0^{r*}, b\rangle ~ {\tt be} ~  
\mathsf{CR\mbox{-}eq}(M_0 =_\beta^k \cdot) ~ 
  {\tt in} ~ \langle a, M_0^{r*}, b+1\rangle$
  
\item $\mathsf{CR\mbox{-}eq}(M_0 =_\beta^{k} \cdot \to \cdot) = $ 
 ${\tt let} ~ \langle a, M_0^{r*},b\rangle ~ {\tt be} ~ 
\mathsf{CR\mbox{-}eq}(M_0 =_\beta^k \cdot)
   ~ {\tt in} ~ 
 \displaystyle{ \langle a+ \frac{1}{2}\mathbf{2}_r^{\vert M_0\vert}, 
   M_0^{(r+1)*}, 
       \frac{1}{2}\mathsf{M_{}} + 2^{\mathsf{M}^{2^{b}}}
     \rangle}$
  
\end{enumerate}
\end{definition}
Note that in the definition of $\mathsf{CR\mbox{-}eq}$, as shown by the use of 
$\cdot$, we use no information on $N$ such that $M_0=_\beta N$, 
but only by the use of the length of $=_\beta$ and case analysis of 
$\to$ or $\leftarrow$. 
From Definition \ref{def:TermSize} and Proposition \ref{prop:length}, 
$\mathsf{TermSize}(M_0=_\beta M_k)$ is well-defined by induction on $=_\beta$. 
From the definition above, $\mathsf{CR\mbox{-}eq}$ is also a function in 
the fourth level of the Grzegorczyk hierarchy (non-elementary).    

\begin{proposition}[Church-Rosser for $\beta$-equality]
If $M_0 =_\beta^k M_k$ with length $k = l+r$ 
where $l = \sharp l[0,k]$ and $r =\sharp r[0,k]$, 
then we have $\mathsf{CR\mbox{-}eq}(M_0=_\beta^k M_k) 
= \langle m, M_0^{r*}, n\rangle$
such that 
$M_0\twoheadrightarrow^{a} M_0^{r*}$ and 
$M_0^{r*}\twoheadleftarrow^{b} M_k$ with $a\leq m$ and $b\leq n$. 
\end{proposition}
%
%
{\it Proof.} 
By induction on the length of $=_\beta^{(l+r)}$. 
The outline of the proof is the same as that of Lemma \ref{th:main}.
\begin{enumerate}
\item Base cases of $k=1$: 
\begin{itemize}
\item $\mathsf{CR\mbox{-}eq}(M_0 \leftarrow \cdot) 
= \langle 0, M_0^{0*}, 1\rangle$: 

We have $M_0 \equiv M_0^{0*} \leftarrow M_1$ for some $M_1$.

\item $\mathsf{CR\mbox{-}eq}(M_0\rightarrow \cdot) 
= \langle \frac{1}{2}\vert M_0\vert, M_0^{*}, 
\frac{1}{2}\vert M_0\vert^2\rangle$:

We have $M_0 \to M_1$ for some $M_1$, 
and then $M_0 \twoheadrightarrow^a M_0^{*}$ 
with $a\leq \frac{1}{2}\vert M_0\vert$ 
and $M_0^{*} \twoheadleftarrow^b M_1$ 
with $b\leq \mathsf{Rev}(\vert M_0\vert , 1) = \frac{1}{2}\vert M_0\vert^2$.

\end{itemize}
\item Step cases:
\begin{itemize}
\item $\mathsf{CR\mbox{-}eq}(M_0=_\beta^{k} \cdot \leftarrow \cdot)  = $
${\tt let} ~ \langle a, M_0^{r*}, b\rangle ~ {\tt be} ~  
\mathsf{CR\mbox{-}eq}(M_0 =_\beta^k \cdot) ~ 
  {\tt in} ~ \langle a, M_0^{r*}, b+1\rangle$: 

From the induction hypothesis, we have 
$M_0 \twoheadrightarrow^m M_0^{r*}$ with $m\leq a$
and $M_0^{r*} \twoheadleftarrow^n M_2 \leftarrow M_3$ for some $M_2, M_3$
with $n \leq b$. 
Then we have the same common reduct $M_0^{r*}$ and 
$n+1 \leq b+1$ from $M_0^{r*} \twoheadleftarrow^{n+1} M_3$.
  
\item $\mathsf{CR\mbox{-}eq}(M_0=_\beta^{k}\!\!\cdot\!\!\to\!\!\cdot)$ $=$ 
 ${\tt let} ~ \langle a, M_0^{r*},b\rangle ~ {\tt be} ~ 
\mathsf{CR\mbox{-}eq}(M_0\!=_\beta^k\!\!\cdot)
   ~ {\tt in} ~ 
 { \langle a+ \frac{1}{2}\mathbf{2}_r^{\vert M_0\vert}, 
   M_0^{(r+1)*}, 
       \frac{1}{2}\mathsf{M} + 2^{\mathsf{M}^{2^{b}}}
     \rangle}$: 

From the induction hypothesis, we have 
$M_0\twoheadrightarrow^m M_0^{r*}$ with $m\leq a$
and $M_0^{r*} \twoheadleftarrow^n M_2\to M_3$ for some $M_2,M_3$ with $n\leq b$.
We also have $M_2^*\twoheadleftarrow^c M_3$ 
with $c\leq \frac{1}{2}\vert M_2\vert \leq \frac{1}{2}\mathsf{M}$, 
and then $M_0^{(r+1)*} \twoheadleftarrow^d M_2^*$ where 
$$
d ~~ \leq ~~ \mathsf{Mon}(\vert M_2\vert, b, 1)
 ~~ \leq ~~ \mathsf{Mon}(\mathsf{M_{}}, b, 1) ~~ = ~~ 2^{\mathsf{M}^{2^b}}.
$$
Hence, we have a common reduct $M_0^{(r+1)*}$ such that 
$M_0\twoheadrightarrow^m M_0^{r*} \twoheadrightarrow^e M_0^{(r+1)*}$ 
where
$$
\hspace{40mm}
 m + e ~~ \leq ~~  a + \frac{1}{2}\vert M_0^{r*}\vert
 ~~ \leq ~~  a + \frac{1}{2}\mathbf{2}_r^{\vert M_0\vert}.
\hspace{40mm} \Box
$$
\end{itemize}
\end{enumerate}
\begin{example}
The Church numerals $\mathbf{c}_n= \lambda f x. f^n(x)$ 
are defined as usual due to Rosser \cite{Barendregt1984}, 
where we write $F^0 (M) = M$, and $F^{n+1}(M) = F(F^n (M))$.
We define $N_i$ such that 
$N_1 = \mathbf{c}_2$, and 
$N_{n+1} = N_n \mathbf{c}_2$.
We also define $M_1 = \mathbf{c}_1 p (N_n  p q)$
and $M_2 = N_n p (\mathbf{c}_1 p q)$ 
with fresh variables $p$ and $q$ for $n\geq 4$.  
We might have $M_1 =_\beta M_2$, but the length of $=_\beta$ is not trivial.
From the fact that 
$N_n \twoheadrightarrow^a \lambda f\lambda x. f^{\mathbf{2}_n^1}(x)$
with $a\leq \mathbf{2}_n^1$,
indeed we prove $M_1 =_\beta M_2$ as follows: 

$M_1 
\twoheadrightarrow 
\mathbf{c}_1 p ((\lambda f\lambda x. f^{\mathbf{2}_n^1}(x)) p q)
\twoheadrightarrow^2 
\mathbf{c}_1 p (p^{\mathbf{2}_n^1}(q))
\twoheadrightarrow^2 
p (p^{\mathbf{2}_n^1}(q))$, and similarly 
$p^{\mathbf{2}_n^1}(p(q))
\twoheadleftarrow M_2$. 
\\
Hence, the length of $=_\beta$ is at most $2\times(4+\mathbf{2}_n^1)$, and 
the size of the common reduct is $1+2\times(\mathbf{2}_{n+1}^1+1)$,
although $\vert M_1\vert = \vert M_2\vert = 8n+1$.
The example suggests that there is plenty of room for improvement 
of the upper bound.  
Note that $M_1\twoheadrightarrow 
p^{\mathbf{2}_n^1+1}(q) \twoheadleftarrow M_2$ is regarded as a base case 
in the sense of Example 1.
\end{example}

\section{Concluding remarks and further work}

The main lemma revealed that a common contractum $P$
from $M_0$ and $M_k$ with $M_0 =_\beta^k M_k$ can be determined by 
(i) $M_0$ and the number of occurrences of $\to$ in $=_\beta$, and also 
by 
(ii) $M_k$ and that of $\leftarrow$. 
In general, we have $2^k$ patterns of reduction graph for $=_\beta^k$ 
as a combination of $\to$ and $\leftarrow$ with length $k$. 
This lemma means that $2^k$ patterns of graph can be grouped 
into $(k+1)$ classes with 
$\displaystyle{_k C_i}$ patterns $(i=0, \ldots, k)$, 
like Pascal's triangle. 
As demonstrated by Example 1, we have common contractums 
$\langle M_0^{(k-i)*}, M_k^{i*}\rangle$ for each class $(i=0, \ldots, k)$, 
contrary to an exponential size of the patterns of reduction graph.
Moreover, Corollary \ref{cor:RefinedTheorem} 
provides an optimum common contractum $M_r^{m_l*}$ for $M_0=_\beta^k M_k$ 
in terms of Takahashi translation, which is one of important consequences of 
the main lemma.

The main lemma depends only on 
Proposition \ref{prop:Mono} 
and Lemma \ref{lem:WeakCR}, 
which can be expounded geometrically as parallel and flipped properties
respectively. Hence, if there exists an arbitrary reduction strategy $*$ that 
satisfies both properties, 
then the main lemma can be established.
In fact, the main lemma holds even for $\beta\eta$-equality,  
because for $\beta\eta$-reduction, under an inside-out development 
we still have Lemma \ref{lem:WeakCR}, 
Proposition \ref{prop:Mono}, 
and Proposition \ref{prop:CompleteDV} without bounds 
as observed already in \cite{KMY2014}. 
This implies that 
under a general framework with such a strategy,   
it is possible to analyze quantitative properties of rewriting systems
in the exactly same way, 
and indeed 
$\lambda$-calculus with $\beta\eta$-reduction and 
weakly orthogonal higher-order rewriting systems \cite{vanOostrom1999,DVO2008}
are instances of these systems. 
Moreover, this general approach is available as well 
for compositional Z \cite{NF2016} that is 
an extension of the so-called Z property \cite{DVO2008} 
(property of a reduction strategy that is cofinal and monotonic),   
which makes it possible to apply a divide and conquer method for proving 
confluence.

In order to analyze reduction length of the Church-Rosser theorem, 
we provided measure functions $\mathsf{Len}$, $\mathsf{TermSize}$, 
$\mathsf{Mon}$, and $\mathsf{Rev}$. 
In terms of the measure functions, bound functions are obtained 
for the theorem for $\beta$-reduction and $\beta$-equality,  
explicitly together with common contractums.
A bound on the valley size for the theorem for $\beta$-equality is 
obtained by induction on the length of $=_\beta$.
Compared with \cite{KS2013}, 
the use of $\mathsf{TermSize}$ is important 
to set bounds to the size of terms, in particular, 
for the theorem for $\beta$-equality. 
Given $M=_\beta N$, then there exists some constant 
$\mathsf{TermSize}(M=_\beta N)$, 
and under the constant bound functions can be provided by 
induction only on the length of $=_\beta$ with neither information on 
$M$ nor $N$, including the size of a common contractum.

In addition, 
based on Corollary \ref{cor:RefinedTheorem}, 
it is also possible to analyze the valley size of $M_0=_\beta^{(l+r)} M_{l+r}$ 
in terms of $M_r^{m_l*}$: 
In the base case of $m_l =0$, 
the valley size is bounded simply by $l$ and $r$, for instance, see Example 2; 
in the maximum case of $m_l=\min\{l,r\}$, 
the valley size is at most that of the theorem for $\beta$-reduction 
as observed in Example 1; and this analysis will be discussed elsewhere. 

Towards a tight bound, 
our bound depends essentially 
on Proposition \ref{prop:CompleteDV} 
and Lemma \ref{lemma:sizeofstar}. 
Proposition \ref{prop:CompleteDV} provides an optimal reduction,
since we adopted the so-called minimal complete development 
\cite{HS2008,Khasidashvili1988,Sorensen2007}.
For the bound on the size of $M^*$, 
Lemma \ref{lemma:sizeofstar} 
can be proved, in general, 
under some function $f(x)$ such that $f(x) \times f(y) \leq f(x+y)$,
which may lead to 
a non-elementary recursive function, as described by $\mathsf{Len}$.

\subparagraph*{Acknowledgements}
The author is grateful to Roger Hindley 
for his valuable comments on this work,  
Pawel Urzyczyn 
for his interest in the new proof, 
Aart Middeldorp and Yokouchi Hirofumi for constructive discussions, 
and the anonymous referees and the editors for useful comments.
This work was partially supported by JSPS KAKENHI Grant Number JP25400192.

\providecommand{\urlalt}[2]{\href{#1}{#2}} 
\providecommand{\doi}[1]{doi:\urlalt{http://dx.doi.org/#1}{#1}}

\end{document}